\documentclass[aps,pra,twocolumn,showpacs,floatfix]{revtex4}
\usepackage{epsfig}
\usepackage{graphicx}
\usepackage{dcolumn}
\begin{document}
\title{Blackbody radiation shift in a $^{43}$Ca$^{\rm{+}}$ ion optical frequency standard }
\author{Bindiya Arora}
\affiliation{Department of Physics and Astronomy, University of Delaware, Newark, Delaware 19716-2593}
\author{M. S. Safronova}
\affiliation{Department of Physics and Astronomy, University of Delaware, Newark, Delaware 19716-2593}
\author{Charles W. Clark}
 \affiliation{
Physics Laboratory,
National Institute of Standards and Technology,
Technology Administration,
U.S. Department of Commerce, 
Gaithersburg, Maryland 20899-8410}
 
\begin{abstract}
Motivated by the prospect of an optical 
frequency standard based on $^{43}$Ca$^+$, we calculate the 
blackbody radiation (BBR) shift of the  $4s_{1/2}-3d_{5/2}$ clock transition, which is a major component of the uncertainty budget. 
The calculations are based on the relativistic all-order single-double 
method where all single and double excitations of the Dirac-Fock 
wave function are included to all orders of perturbation theory.  
Additional calculations are conducted for the dominant contributions 
in order to evaluate some omitted high-order corrections and estimate 
the uncertainties of the final results.
The BBR shift obtained for this transition is $0.38(1)$ Hz.  
The  tensor polarizability of the $3d_{5/2}$ level is 
also calculated and its uncertainty is evaluated as well. 
Our results are compared with other calculations.  
\date{\today}
\end{abstract}
\maketitle

The definition for the International System of Units (SI) of time, the second, is based on the 
microwave transition between the two hyperfine levels of 
the ground state of  $^{133}$Cs. 
The advancements in experimental techniques such as laser frequency 
stabilization, atomic cooling and trapping, etc. have made the realization
of the SI unit of time possible  to 15 digits. 
The operation of atomic clocks is generally carried out 
at room temperature, whereas the definition 
of the second refers to the clock transition in an atom at absolute zero. 
This implies that the clock transition frequency should be corrected in practice for the 
effect of finite temperature of which the leading contributor is the blackbody radiation (BBR) shift. 
A number of groups~\cite{angstmann,beloy} have accurately 
calculated the BBR shift at room temperature affecting 
the Cs microwave frequency standard. 
A significant further improvement in frequency standards 
is possible with the use of optical transitions. 
The frequencies of feasible optical clock transitions are five orders 
of magnitude larger than the relevant microwave transition frequencies, thus making it theoretically
possible to reach relative uncertainty of $10^{-18}$.
The ability to develop more precise optical frequency 
standards will open ways to improve Global Positioning System
(GPS) measurements and tracking of deep-space probes, perform more accurate 
measurements of the physical constants and tests of
fundamental physics such as searches for nonlinearity of quantum mechanics, 
gravitational waves, etc. 

One of the promising schemes for the optical 
frequency standard with a single $^{43}$Ca$^+$ 
ion trapped in a Paul trap was proposed   by Champenois \textit{et al.}~\cite{champ}. 
A major advantage of using $^{43}$Ca$^+$ ions is that the radiation 
required for cooling, repumping, and clock transition is easily 
produced by non-bulky solid state or diode lasers.
In this work, we calculate the shift in a  $^{43}$Ca$^+$ 
optical clock transition due to blackbody radiation. 
The clock transition involved is $4s_{1/2} \rightarrow 3d_{5/2}\mbox{ }F=6\mbox{ } m_F=0$.  

The accuracy of optical frequency standards is  
limited by the frequency shift in the clock transitions caused by 
the interaction of the ion with external fields. 
One such shift in the atomic levels is caused by external magnetic fields 
(Zeeman shift). 
The first-order Zeeman shift can be eliminated by using 
isotopes with half-integer nuclear spin~\cite{champ}. 
The second-order Zeeman shift depends on the choice 
of hyperfine sublevels involved in the optical transition. 
The particular transition scheme used to minimize second-order Zeeman 
shift in $^{43}$Ca$^+$ is $4s_{1/2}{\rightarrow}3d_{5/2}\mbox{ }F=6\mbox{ }m_F=0$. 
The effect of this shift on the clock transition is of the order of -0.09(9) Hz at 0.1 $\mu$T at room temperature~\cite{champ}. 
There are some additional shifts in the clock transition frequency 
due to the second-order  Doppler effect. These are on the order of mHz for ions at room temperature~\cite{champ,kajita1}.
There have been a number of accurate calculations 
of the Zeeman and Doppler shifts~\cite{champ,kajita1}; however, there has been 
no accurate study of the BBR shift in this optical transition. 
As we show here, the BBR shift is the major component in the uncertainty budget of this frequency standard at room temperature.

The frequency-dependent electric field $E$ radiated by a black body at  
temperature T is given by the Planck's radiation law. 
The frequency shift of an ionic state due to such an electric field can be related to the state's static scalar polarizability $\alpha_0$  by 
	\begin{equation}
		\Delta\nu = -\frac{1}{2} (831.9\mbox{ }\rm{V/m})^2\left(\frac{\rm{T(K)}}{300}\right)^4 \alpha_0 (1+\eta)
		\label{eq-st},
	\end{equation} 
where $\eta$ is a small ``dynamic'' correction~\cite{sergbbr}. 
We estimate that this dynamic correction is negligible with comparison to the present 3 \% accuracy of our calculations.
The effect of the vector and tensor parts of the polarizability average out due to the isotropic nature of the electric field radiated by the black body.
However, for completeness, we also present results for the static tensor 
polarizability of the $3d_{5/2}$ state.

The overall BBR shift of the clock transition frequency is calculated as the difference between 
the BBR shifts of the individual levels involved in the transition: 
	\begin{eqnarray}
		\Delta_{\rm{BBR}}(4s\rightarrow 3d_{5/2}) &=& -\frac{1}{2}\left(\alpha_{0}(3d_{5/2})-\alpha_{0} 
(4s_{1/2})\right)\nonumber\\
		&\rm{x}&(831.9~\textrm{V/m})^2\left(\frac{\rm{T(K)}}{300}\right)^4.\label{eq-formula}
	\end{eqnarray}
Therefore, the evaluation of the BBR shift involves accurate calculation of 
static scalar polarizabilities for the 4$s_{1/2}$ ground and 3$d_{5/2}$ excited states. 
We use the relativistic single-double (SD) all-order method to find the electric-dipole matrix 
elements used in the calculation of dominant contributions to the polarizability values. 
The relativistic all-order method is one of the most accurate 
methods currently being used in atomic structure calculations. 
We refer the reader to  Refs.~\cite{liao,relsd} for the detailed description of 
this approach.
\begin{table}
\caption{\label{cas}Contributions to the $4s_{1/2}$ scalar ($\alpha_{0}$) static polarizabilities in $^{43}$Ca$^+$ and 
their uncertainties in units of $a_0^3$. The values of corresponding E1 matrix elements are given in $ea_0$.}
\begin{ruledtabular}
\begin{tabular}{ccc}
\multicolumn{1}{c}{Contribution} &
\multicolumn{1}{c}{$\langle k\|D\|4s_{1/2}\rangle$} &
\multicolumn{1}{c}{$\alpha_0$}\\
\hline 
$4s_{1/2}-4p_{1/2}$ & 2.898 &   24.4(5)  \\   
$4s_{1/2}-5p_{1/2}$ & 0.076 &    0.007 \\
$4s_{1/2}-6p_{1/2}$ & 0.085 &    0.007 \\
$4s_{1/2}-4p_{3/2}$ & 4.099 &   48.4(1.0)  \\
$4s_{1/2}-5p_{3/2}$ & 0.091 &    0.010  \\
$4s_{1/2}-6p_{3/2}$ & 0.112 &    0.012  \\
$\alpha_{\rm{core}}$  &       &    3.25(17)  \\
$\alpha_{\rm{tail}}$  &       &    0.006(6)  \\
$\alpha_{\rm{total}}$ &       &    76.1(1.1)         
\end{tabular}   
\end{ruledtabular}
\end{table}

The calculation of the static scalar polarizability of 
an atom can be separated into the calculations of the
ionic core contribution $\alpha_{\rm{core}}$ and a valence contribution. 
The static ionic core polarizability values were calculated using random-phase 
approximation (RPA) in Ref.~\cite{datatab2}. 
The valence scalar and tensor polarizabilities of an atom in a state 
$v$ are expressed as the sums over all intermediate states $k$ allowed by 
the electric-dipole selection rules:
	\begin{eqnarray}
		\alpha_{0} &=& \frac{2}{3(2j_v+1)}\sum_k \frac{\left\langle 
k\left\|D\right\|v\right\rangle^2}{E_k-E_v},\label{eq-sca} \\
         \alpha_{2}&=&-4C\sum_k(-1)^{j_v+j_k+1}
         \left\{   
         \begin{array}{ccc}
         j_v & 1 & j_k \\
         1 & j_v & 2 \\
         \end{array}
         \right\} 
         \frac{{\left\langle 
          k\left\|D\right\|v\right\rangle}^2}{ 
         E_k-E_v}, \label{eq-ten}\nonumber \\ 
        C &=&  
       \left(\frac{5j_v(2j_v-1)}{6(j_v+1)(2j_v+1)(2j_v+3)}\right)^{1/2}, 
     \end{eqnarray}    
where $\left\langle k\left\|D\right\|v\right\rangle$ are the reduced electric-dipole (E1) matrix elements
and $E_i$ is the energy of a state $i$. In these equations, and hereafter, we use the conventional
system of atomic units, a.u., in which $e, m_{\rm e}$, $4\pi 
\epsilon_0$ and the reduced Planck constant $\hbar$ have the 
numerical value 1.  Polarizability in a.u. has the dimensions of 
volume, and its numerical values presented here are thus expressed
in units of $a^3_0$, where $a_0\approx0.052918$~nm is the Bohr radius.
The atomic units for $\alpha$ can be converted to SI units via
 $\alpha/h$~[Hz/(V/m)$^2$]=2.48832$\times10^{-8}\alpha$~[a.u.], where
 the conversion coefficient is $4\pi \epsilon_0 a^3_0/h$ and 
 Planck constant $h$ is factored out. 
Experimental energies from Ref.~\cite{NIST1} have been used for the
dominant contributions to the polarizability. 

We use the B-spline method to reduce infinite sums in Eqs.~(\ref{eq-sca})
 and~(\ref{eq-ten}) to a finite 
number of terms~\cite{bspline} with 70 splines 
of order 11 for each angular momentum constrained to 
a spherical cavity with $R = 220$~a.u. The particular choice of such a large cavity 
ensures that all valence orbitals of interest   
are enclosed. In our case, the relevant excited states up to
$12d$ states fit inside this cavity.
 We also separate the polarizability calculation into two parts,
 the main term $\alpha_{\rm{main}}$ containing the first few contributions from the states that fit inside the cavity, with the remaining part designated by $\alpha_{\rm{tail}}$. The main contribution is calculated using the all-order matrix elements obtained in the  present work and experimental energies \cite{NIST1}. The tail contribution is evaluated in the 
 Dirac-Fock (DF) approximation. We included as many states as practical in  $\alpha_{\rm{main}}$ for the  $3d_{5/2}$ state in order to reduce the uncertainty in the remainder. 
\begin{table}
\caption{\label{cap}Contributions to the $3d_{5/2}$ scalar ($\alpha_{0}$) and tensor ($\alpha_{2}$) static polarizabilities 
in $^{43}$Ca$^+$ and their uncertainties in units of $a_0^3$. The values of corresponding E1 matrix elements are  given 
in $ea_0$.}
\begin{ruledtabular}
\begin{tabular}{cccc}
\multicolumn{1}{c}{Contribution} &
\multicolumn{1}{c}{$\langle k\|D\|3d_{5/2}\rangle$} &
\multicolumn{1}{c}{$\alpha_0$}&
\multicolumn{1}{c}{$\alpha_2$}\\
\hline
$3d_{5/2}-4p_{3/2}$  &  3.306  &   22.78(25)&-22.78(25) \\
$3d_{5/2}-5p_{3/2}$  &  0.148  &   0.011(2) & -0.011(2) \\
$3d_{5/2}-6p_{3/2}$  &  0.095  &   0.004    & -0.004    \\[0.5pc]
$3d_{5/2}-4f_{5/2}$  & 0.516 &  0.120(3) &  0.137(3)  \\
$3d_{5/2}-5f_{5/2}$  & 0.319 &  0.039(2) &  0.044(2)  \\
$3d_{5/2}-6f_{5/2}$  & 0.224 &  0.018(1) &  0.020(1)  \\
$3d_{5/2}-7f_{5/2}$  & 0.169 &  0.010    &  0.011     \\
$3d_{5/2}-8f_{5/2}$  & 0.134 &  0.006    &  0.007     \\
$3d_{5/2}-9f_{5/2}$  & 0.110 &  0.004    &  0.004     \\
$3d_{5/2}-10f_{5/2}$ & 0.093 &  0.003    &  0.003     \\
$3d_{5/2}-11f_{5/2}$ & 0.079 &  0.002    &  0.002     \\
$3d_{5/2}-12f_{5/2}$ & 0.069 &  0.002    &  0.002     \\[0.5pc]
$3d_{5/2}-4f_{7/2}$  & 2.309 &  2.392(53) & -0.854(19) \\
$3d_{5/2}-5f_{7/2}$  & 1.428 &  0.773(33) & -0.276(12)  \\
$3d_{5/2}-6f_{7/2}$  & 1.000 &  0.350(12) & -0.125(4)  \\
$3d_{5/2}-7f_{7/2}$  & 0.757 &  0.191(7)  & -0.068(3)  \\
$3d_{5/2}-8f_{7/2}$  & 0.600 &  0.117(4)  & -0.042(2)  \\
$3d_{5/2}-9f_{7/2}$  & 0.492 &  0.077(3)  & -0.028(1)  \\
$3d_{5/2}-10f_{7/2}$ & 0.414 &  0.054(2)  & -0.019(1)  \\
$3d_{5/2}-11f_{7/2}$ & 0.354 &  0.039(1)  & -0.014(1)  \\
$3d_{5/2}-12f_{7/2}$ & 0.308 &  0.029(1)  & -0.011    \\[0.5pc]
$\alpha_{\rm{core}}$   &       &   3.25(17) &                         \\
$\alpha_{\rm{tail}}$   &       &   1.7(1.1) &       -0.5(3)     \\
$\alpha_{\rm{total}}$  &       &   32.0(1.1)&         -24.5(4)    
\end{tabular}  
\end{ruledtabular}
\end{table}

Table~\ref{cas} shows the contributions 
from the individual transitions to the $\alpha_0$
of the 4$s_{1/2}$ ground state of $^{43}$Ca$^+$.
The main contributions are listed separately along with the 
respective values of the electric-dipole matrix elements. The tail contributions are grouped together as $\alpha_{\rm{tail}}$. 
The total contribution of the
$4s_{1/2}-4p_{1/2}$ and $4s_{1/2}-4p_{3/2}$ transitions is overwhelmingly dominant.   
We use our \textit{ab initio} SD all-order value for these matrix elements 
based on the comparison of the all-order results for alkali-metal atoms
with experiment \cite{relsd} and assign these values 1 \% uncertainty. 

The lifetimes of the $4p_{1/2}$ and $4p_{3/2}$ states have been measured by 
Jin and Church \cite{jin} using a variant of the collinear laser-beam - ion-beam
spectroscopy technique in 1993. The experimental results, $\tau(4p_{1/2})=7.098(20)$~ns
and $\tau(4p_{3/2})=6.924(19)$~ns \cite{jin} were found to be in significant 
disagreement with perturbation theory calculations at that time. By 1993, similar
discrepancies were found between other measurements carried out using the 
same technique and theoretical calculations of light alkali-metal atom lifetimes. 
These disagreements in alkali-metal atoms were resolved a few years later 
when new measurements became available (see \cite{ADNDT,relsd} and references therein).
 In all cases, new experimental results were found to be in excellent agreement 
 with precise theoretical calculations. Recently, excellent agreement was also 
 found between experimental and theoretical calculations of the $3d$ lifetimes in Ca$^{+}$~\cite{usca}.
 Just as in the case of earlier comparison of theory and experiment for alkali-metal atoms,
 we find significant discrepancy between out present calculations yielding 
 $\tau(4p_{1/2})=6.875$~ns, $\tau(4p_{3/2})=6.686$~ns and measurements of Ref.~\cite{jin}.
 It is interesting to make a direct comparison of the $4p_j-4s$ theoretical and experimental 
 matrix elements. The contributions of the $4p-3d$ transitions into the $4p$ lifetimes
 are small, as shown by comparison of our theory values for the Einstein A-coefficients $A_{ab}$
 for the transitions to the $4s$ state, $A_{4p_{1/2} 4s}=136.0$~MHz and $A_{4p_{3/2} 4s}=139.7$~MHz,
 and for the transitions to the $3d_j$ states,  $A_{4p_{1/2} 3d_{3/2}}=9.452$~MHz,
 $A_{4p_{3/2} 3d_{3/2}}=0.997$~MHz, and $A_{4p_{3/2} 3d_{5/2}}=8.877$~MHz. Substituting the 
 theoretical values for the  $A_{ab}$, $b=3d_{3/2}$, $3d_{5/2}$, and experimental $4p_j$ lifetimes
 from \cite{jin} into the formula $\tau_a=1/\sum_b A_{ab}$, we obtain the following values for the $4p_j-4s$
 reduced electric-dipole matrix elements: $\langle 4p_{1/2}\|d\|4s\rangle^{expt}=2.849(4)$ 
 and $\langle 4p_{1/2}\|d\|4s\rangle^{expt}=4.023(6)$. While the uncertainties 
 of these values include the uncertainties in the theoretical values of the 
 $4p-3d$ matrix elements, their contributions are negligible. 
 Our theoretical values differ from these experimental matrix elements by 1.7\% and 1.9\% for
 the $4p_{1/2}-4s$ and  $4p_{3/2}-4s$ transitions, respectively. Our theoretical values for these
 transitions in K agree with experiment values within experimental uncertainties (0.13\%).
 We note some differences between the contributions of the various correlation correction terms
 in the K and Ca$^+$ calculations of these matrix elements.   
  Therefore, it would very interesting to 
 see a new measurement of these matrix elements in Ca$^+$ to determine if the discrepancies between 
 the theoretical and experimental values of the $4p$ lifetimes will be resolved as in the case of the
 alkali-metal atoms.

 Our all-order SD
values are used for other main term contributions as well but their contribution 
is negligible. The uncertainty of the core contribution is taken to be 
5 \% based on the comparison of the RPA approximation with experimental values for noble gases. 
\begin{table}
\caption{\label{comp1} Comparison of static scalar polarizabilities for
 $4s_{1/2}$ and $3d_{5/2}$ states and  blackbody radiation shift for the $4s_{1/2}-3d_{5/2}$ transition of 
$^{43}$Ca$^{\rm{+}}$ ion  at T=300 K. The polarizability values
 are in $a_0^3$ and BBR shift is in Hz.}
\begin{ruledtabular}
\begin{tabular}{ccccc}
\multicolumn{1}{c}{} &
\multicolumn{1}{c}{Present work} &
\multicolumn{1}{c}{Ref.~\cite{champ}}&
\multicolumn{1}{c}{Ref.~\cite{kajita1}} &
\multicolumn{1}{c}{Ref.~\cite{theo1}} \\
\hline
$\alpha_0(4s_{1/2})$  &  76.1(1.1) &  76  &  73 & 70.89(15) \\
$\alpha_0(3d_{5/2})$  &  32.0(1.1) &  31  &  23 &          \\

BBR shift             &  0.38(1)   &  0.39(27)& 0.4 
\end{tabular}   
\end{ruledtabular}
\end{table}

Table~\ref{cap} shows the detailed breakdown 
of the contributions to the 3$d_{5/2}$ state  polarizabilities. 
In this case, the allowed transitions
are to  $np_{3/2}$ and $nf_{5/2,7/2}$ states. 
The two dominant contributions come from the $3d_{5/2}-4p_{3/2}$ and $3d_{5/2}-4f_{7/2}$ transitions. 
The sum of the contribution from the $3d_{5/2}-nf$ transitions
 converges more slowly than the $ns-np$ sums; therefore, 
 we include in the main term the nine lowest $3d_{5/2}-nf_j$ transitions for each $j$.  
In addition to the SD all-order calculations, we carry out semi-empirical scaling procedure 
for the two lowest 3$d_{5/2}-np_{3/2}$ transitions and the 
three lowest 3$d_{5/2}-nf$ transitions to evaluate some classes 
of the correlation corrections omitted by the current all-order calculation.

We also conduct additional more complicated \textit{ab initio} calculations that partially include
triple excitation (SDpT) since those appear to be important for these types of transitions. 
The scaling procedure and SDpT calculations are described in Refs.~\cite{relsd,1,usca}.
We conducted a more detailed study of the uncertainties in the values of the final matrix elements for the first 
transition in each sum, i.e.
$3d_{5/2}-4p_{3/2}$, $3d_{5/2}-4f_{5/2}$, and $3d_{5/2}-4f_{7/2}$. They are determined as the 
maximum difference between the final results and the \textit{ab initio} and scaled SDpT values. 
The remaining uncertainties for the $3d_{5/2}-nf$ contributions are determined as the differences
 between the final values (either scaled SD or SDpT) and the \textit{ab initio} SD values.   
The all-order calculation of such a large number of matrix 
elements and evaluation of their uncertainties makes the calculation of the 3$d_{5/2}$ state polarizability 
much more challenging than the calculation of the ground state polarizability. 

The tail contribution of the 3$d_{5/2}$ state is almost 5\% 
of the total polarizability. 
Such a large tail leads to a large uncertainty in the total polarizability. 
In order to evaluate the uncertainty in the tail, we calculated the last few 
main terms using the DF approximation and compared the results with our all-order values. 
We found that the DF approximation overestimates the polarizability contributions 
by about 60 \% relative to the all-order value. 
To improve our accuracy, we scaled the tail contribution with this ratio and 
took the difference of the DF and scaled DF result to be the uncertainty of
the tail value. The uncertainties from  all terms are added in quadrature to obtain the 
uncertainties of the final polarizability values. 

We also list the contributions from various transitions to the $3d_{5/2}$ tensor polarizability in Table~\ref{cap}. 
The $3d_{5/2}-4p_{3/2}$  transition also
gives the dominant contribution to the tensor polarizability. 
The tail contribution is less important for $\alpha_2$ 
and the uncertainty in $\alpha_2$ calculation is less than that of $\alpha_0$. 

We use the scalar polarizability values to evaluate the
shift of a clock transition due to blackbody radiation at 300 K.  
The final shift in the clock transition for a $^{43}$Ca$^+$ ion is found to be (0.380 $\pm$ 0.013) Hz. 
The overall uncertainty of the result is dominated by the contribution of
highly-excited $nf_{7/2}$ states to $3d_{5/2}$ polarizability.

Black body radiation also shifts the hyperfine splittings of the ground and excited states \cite{beloy,angstmann} 
	although this is third-order effect  and thus expected to be small.
	For completeness, we have calculated this effect which is proportional to the difference in the atomic polarizabilities between the two ground hyperfine structure states F of the atom. This is exactly the sort of calculation that is involved in the determination of the BBR shift in the microwave standard. 
	The F-dependent polarizability can be written as the sum of three terms,
	each involving a reduced hyperfine matrix element and two reduced electric-dipole
	matrix elements \cite{beloy,angstmann}. We estimated the magnitude of the frequency shift of the ground 
	state hyperfine transition using DF values of these reduced matrix elements. For 
	consistency, we used the same set of B-splines.	
	Further details of the method can be found in Refs.~\cite{beloy,angstmann}.
	The frequency shift in the ground-state hyperfine transition 
	of $^{43}$Ca$^+$  at room temperature is calculated to be
	$4.2 \times 10^{-6}$ Hz. 
	The magnitude of this third-order correction is 
	approximately $10^{-5}$ times lower than the 
	second-order BBR shift 
	and hence can be neglected. 
	The effect of electric field
	on the frequency of the  $3d_{5/2}$ state hyperfine transition is even
	smaller and; therefore, is not expected to affect the BBR shift of the clock	transition.

In Table~\ref{comp1}, we compare our polarizability and BBR shift results with 
previous theoretical studies. In the work done by Champenois \textit{et al.}~\cite{champ},
 only first few transitions  were included in the Stark shift calculations. 
The required oscillator strengths were taken from the Kurucz database~\cite{kuruczdb} in 
Ref.~\cite{champ} and  from Ref.~\cite{wiese} in Ref.~\cite{kajita1}. 
Our value for the ground state polarizability differs from that of Theodosiou~\cite{theo1}
in part because of his omission of the core polarizability contribution in \cite{theo1}
and his use of the Jin and Church \cite{jin} values of the $4s-4p$ matrix elements.
 Our all-order values are in good agreement with those predicted 
by Champenois \textit{et al.}~\cite{champ} and Kajita \textit{et al.}~\cite{kajita1}. 
We have significantly improved the accuracy of the BBR shift in comparison to these calculations 
by using all-order values for the dominant terms 
and also by including the tail and core contributions.

We have increased the accuracy of the polarizability 
values as compared to any other previous calculations 
by using a more accurate dipole-matrix element values 
calculated using relativistic all-order single-double method
making the BBR shift value accurate enough to evaluate the 
performance of a Ca$^{+}$ single-ion frequency standard at room temperature. 

This research was performed
under the sponsorship of the US Department of Commerce, National Institute of Standards
and Technology.

\end{document}